\documentclass[aps,prl,twocolumn,groupedaddress,showpacs,showkeys]{revtex4}

\usepackage{graphicx}
\usepackage{dcolumn}
\usepackage{bm}
\usepackage{amssymb}
\usepackage{amsmath}

\usepackage{color}

\newcommand{\alkor}[1]{\textcolor{red}{#1}}

\begin{document}

\title{Synchronization in networks of slightly nonidentical elements}

\author{Alexander~E.~Hramov}
\email{aeh@nonlin.sgu.ru}
\author{Anastasiya~E.~Khramova}
\email{rabbit@nonlin.sgu.ru}
\author{Alexey~A.~Koronovskii}
\email{alkor@nonlin.sgu.ru}
\affiliation{Faculty of Nonlinear
Processes, Saratov State University, Astrakhanskaya, 83, Saratov,
410012, Russia}

\author{S. Boccaletti}
\affiliation{CNR- Istituto dei Sistemi Complessi, Via Madonna del
Piano 10, 50019 Sesto Fiorentino, Floreence, Italy}

\date{\today}

\begin{abstract}
We study synchronization processes in networks of slightly non
identical chaotic systems, for which a complete invariant
synchronization manifold does not rigorously exist. We show and
quantify how a slightly dispersed distribution in parameters can
be properly modelled by a noise term affecting the stability of
the synchronous invariant solution emerging for identical systems
when the parameter is set at the mean value of the original
distribution.
\end{abstract}


\maketitle

Complex networks are the prominent candidates to describe
sophisticated collaborative dynamics in many areas~\cite{revmod}.
Recently, the dynamics of complex networks has been extensively
investigated  with regard to collective (synchronized) behaviors
\cite{booksPhaseSynchr}, with special emphasis on the interplay
between complexity in the overall topology and local dynamical
properties of the coupled units. The usual case considered so far
is that of networks of identical dynamical systems coupled by
means of a complex wiring of connections. In this framework,
several studies have shown how to enhance synchronization
properties, by properly weighting the strengths of the connection
wiring
\cite{Chavez:2005_Networks_PRL,Hwang:2005_Newtwork_PRL,zhoukurths}.

In this paper, we extend the study of synchronization phenomena in
complex networks to the case of slightly non identical coupled
dynamical systems, i.e. networks whose nodes are represented by
dynamical systems each one of them evolving with the same
functional form, but with a different, node dependent, value of
the control parameters. This is motivated by the fact that such a
representation seems a more adequate description of many relevant
phenomena occurring in natural systems, where the hypothesis that
the evolution in different nodes be identical is very often a too
restrictive assumption.

Let us then consider a network of $N$ coupled dynamical systems
with slightly mismatched control parameters described by the
equations
\begin{equation}
\mathbf{\dot
x}_i=\mathbf{F}(\mathbf{x}_i,\mathbf{g}_i)-\sigma\sum\limits_{j=1}^NG_{ij}\mathbf{H}[\mathbf{x}_j],
{\quad}i=1,\dots,N, \label{eq:Network}
\end{equation}
where $\mathbf{x}_i$ are the state vectors in each network node,
$\mathbf{F}$ defines the vector field of the considered systems,
$\mathbf{g}_i$ are the control parameter vectors,
$\mathbf{H}[\cdot]$ is an output function, and $\sigma$ is the
coupling strength. $G$ is the Laplacian matrix of the network. As
so, it is a symmetric zero row sum matrix, it has a real spectrum
of eigenvalues ${\lambda_1\leq\dots\leq\lambda_N}$, $G_{ij}$
($i\neq j$) is equal to 1 whenever node $i$ is connected with node
$j$ and 0 otherwise, and $G_{ii}=-\sum_{j\neq i}G_{ij}$.

When the considered network consists of identical elements (i.e.,
$\mathbf{g}_i=\mathbf{g}$, $\forall i$) the stability of the
synchronous state [$\mathbf{x}_i(t)=\mathbf{x}_s(t)$, $\forall i$]
is known to be determined by the diagonalized linear stability
equation~\cite{Pecora:1998_CS}, yielding $N$ blocks of the form
\begin{equation}
\mathbf{\dot \zeta}_i= \left[
J\mathbf{F}(\mathbf{x}_s,\mathbf{g})-\sigma\lambda_i J
\mathbf{H}(\mathbf{x}_s)\right] \zeta_i, \label{eq:StabNetwork}
\end{equation}
where $J$ is the Jacobian operator. The
blocks~(\ref{eq:StabNetwork}) differ from each other only by the
eigenvalues ${\lambda_1\leq\dots\leq\lambda_N}$ of the coupling
matrix $G$. Replacing $\sigma\lambda_i$ by $\nu$ in
equation~(\ref{eq:StabNetwork}), the behavior of the largest
(conditional) Lyapunov exponent $\Lambda$ vs $\nu$ [also called
master stability function~\cite{Pecora:1998_CS}] completely
accounts for linear stability of the synchronized manifold.
Indeed, the synchronized state associated with $\lambda_1=0$ is
stable when all the remaining blocks related with the other
eigenvalues $\lambda_i$ ($i=2,\dots,N$) of coupling matrix $G$ are
characterized by the negative Lyapunov exponents. So, to analyze
the stability of the synchronized state $\mathbf{x}_s(t)$ in the
network~(\ref{eq:Network}) only one parametric variational
equation
\begin{equation}
\mathbf{\dot \zeta}=\left[
J\mathbf{F}(\mathbf{x}_s,\mathbf{g})-\nu J
\mathbf{H}(\mathbf{x}_s)\right] \zeta
\label{eq:StabNetworkEquation}
\end{equation}
should be considered to obtain the dependence of the master
stability function $\Lambda$ on the parameter $\nu$. Furthermore,
the vector state $\mathbf{x}_s(t)$ may be obtained as a solution
of the uncoupled equation
\begin{equation}
\mathbf{\dot x}_s(t)=\mathbf{F}(\mathbf{x}_s(t),\mathbf{g}).
\label{eq:EvolutionEquation}
\end{equation}

It is worth noticing that the master stability function
$\Lambda(\nu)$ may be negative for a finite interval of
$\nu$-parameter values
$I_{st}=(\nu_1;\nu_2)$~\cite{Pecora:1998_CS} or for an infinite
one ($\nu_2=\infty$), depending on the specific choice of the
functions $\mathbf{F}$ and $\mathbf{H}$. The stability condition
is satisfied if the whole set of eigenvalues $\lambda_i$
($i=2,\dots,N$) multiplied by the same $\sigma$ falls into the
stability interval $I_{st}$, i.e., when conditions
$\sigma\lambda_2>\nu_1$ and $\sigma\lambda_N<\nu_2$ take place
simultaneously. The vector functions $\mathbf{F}(\cdot)$ and
$\mathbf{H}[\cdot]$ are determining the boundaries $\nu_1$ and
$\nu_2$ of the stability interval $I_{st}$, while the eigenvalue
distribution is solely ruled by the topology of the imposed wiring
of connections.

Natural systems, however, are modelled by networks that generally
consist of elements for which parameters might differ. Therefore,
equation~(\ref{eq:StabNetworkEquation}) cannot be seen as a
suitable representation of this case. As soon as the vector
$\mathbf{g}_i$ depends on $i$,  an invariant synchronization
manifold $\mathbf{x}_i(t)=\mathbf{x}_s(t)$, $\forall i$ no longer
exists, and therefore the arguments of the master stability
function do not rigorously apply. However, it has been numerically
verified in Ref. \cite{Chavez:2005_Networks_PRL} that, when the
difference in the parameters is limited to a slight mismatch, the
synchronization behavior keeps on holding in the synchronization
region predicted by the master stability function of the system
corresponding to the parameter vector
${\mathbf{g}=\langle\mathbf{g}_i\rangle}$, where $\langle.\rangle$
stays for the ensemble average on the network nodes.

In the following, we will give ground to such a numerical
evidence, and show that, unless a rigorous treatment of the
complete synchronization state is prevented, an approximate
treatment of these synchronization phenomena is possible under the
assumption of a smallness in the parameter mismatch. Without lack
of generality, we will develop our points with respect to a
subclass of chaotic systems, namely the class of functions
$\mathbf{F}$ describing self sustained chaotic oscillators.

\begin{figure}[tb]
\centerline{\includegraphics*[scale=0.4]{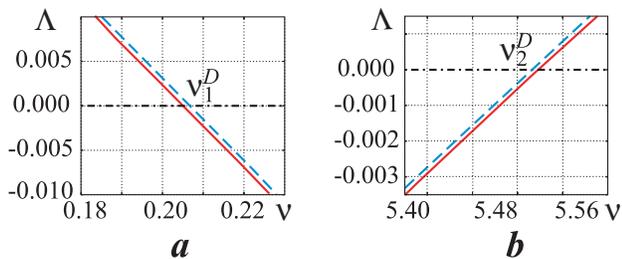}} \caption{(Color
online) The fragments of master stability function corresponding to
those $\nu$ ranges for which $\Lambda(\nu)$ crosses the horizontal
axis. In both plots a) and b), the dashed line refers to
$\Lambda^D(\nu)$ calculated for the network of R\"ossler oscillators
with slightly different parameters $\omega_i$ at $D=3.5$ (see text
for details), whereas the solid line depicts $\Lambda(\nu)$
calculated for the network of identical R\"ossler oscillators with
$\omega_i\equiv\bar{\omega}=1$.} \label{fgr:MSFIDetuned}
\end{figure}

The core idea that justifies our approximation comes from the well
known property~\cite{Pikovsky:2002_SynhroBook} of a pair of
coupled identical chaotic oscillators, where there are two
important values ${\sigma_1<\sigma_2}$ of the coupling strength
$\sigma$ determining the transition to complete synchronization.
Precisely,  $\sigma_1$ determines the blowout bifurcation
\cite{Nagai:1997_BlowoutAndOrbits, ashwin:1635, Lai:1997_BLOWOUT},
when the largest tangential Lyapunov exponent crosses zero. The
second one, $\sigma_2$ corresponds to the loss of the stability in
the tangential direction of the unstable periodic orbits with the
lowest period embedded into the synchronized chaotic manifold.
When the coupling parameter value is in the interval
$\sigma_1<\sigma<\sigma_2$ the bubbling
phenomenon~\cite{Venkataramani:1996_Bubbling, venkataramani:5361}
may be observed. If one considers two coupled \emph{identical}
oscillators the synchronous regime is detected (after expiration
of the transient) for coupling strength values $\sigma>\sigma_1$.
Alternatively, if the control parameters of the coupled
oscillators differ slightly from each other, the synchronous
behavior may be detected for the coupling strength values
exceeding the threshold $\sigma_2$. The same effect takes place if
two identical oscillators in the presence of noise are
considered~\cite{Pikovsky:2002_SynhroBook}. In both cases the
onset of synchronization is shifted towards the larger values of
coupling strength $\sigma$ and determined by the $\sigma_2$-value.

\begin{figure}[tb]
\centerline{\includegraphics*[scale=0.4]{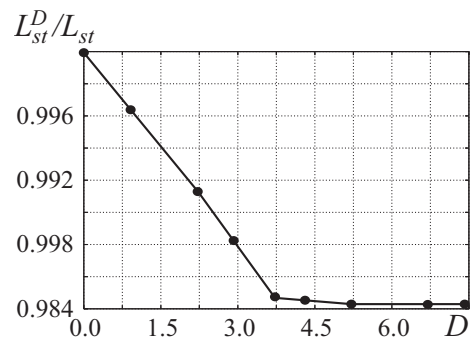}}
\caption{The dependence of the length $L_{st}^D$ of the
stability interval $I_{st}^D$ on the noise intensity $D$ for the
network of R\"ossler oscillators with slightly detuned parameter
$\omega_i$. The value $L^D_{st}$ is normalized on the length
$L_{st}$ of the stability interval $I_{st}$ for the network
consisting of the identical R\"ossler oscillators}
\label{fgr:StabInt}
\end{figure}

The idea here is that the property for synchronization of the
network consisting of elements with slightly mismatched control
parameters $\mathbf{g}_i$ may be estimated by means of the
consideration of the network of identical elements with the
control parameter ${\mathbf{g}=\langle\mathbf{g}_i\rangle}$ in the
presence of noise. In practice, this assumption means that one can
still evaluate the conditional Lyapunov exponents by means of
equation~(\ref{eq:StabNetworkEquation}) at
${\mathbf{g}=\langle\mathbf{g}_i\rangle}$, but the evolution of
the state $\mathbf{x}_s(t)$ around which the conditional exponents
are evaluated has to be taken as a solution of a stochastic
differential equation, i.e. the evolution
equation~(\ref{eq:EvolutionEquation}) has to be replaced by
\begin{equation}
\mathbf{\dot x}_s(t)=\mathbf{F}(\mathbf{x}_s(t))+D\mathbf{\xi}(t),
\label{eq:NoiseEvolutionEquation}
\end{equation}
where $D\mathbf{\xi}(t)$ is a noise term with zero mean value.

\begin{figure}[b]
\centerline{\includegraphics*[scale=0.4]{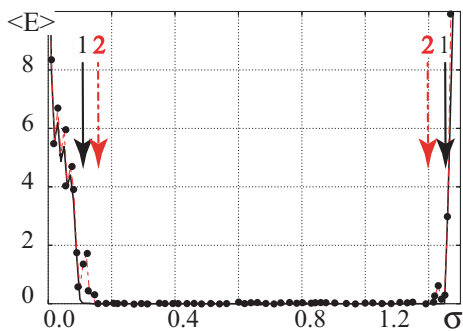}} \caption{(Color
online) $\langle E \rangle$ (see text for definition) vs $\sigma$
for the network of R\"ossler oscillators with identical parameters
$\omega_i\equiv\omega=1$ (black solid line) and for the network of
R\"ossler oscillators with slightly detuned parameters $\Delta
\omega_i=0.07$ (red dashed line with black circles). The arrows
labelled with ``1'' delimit the interval of coupling parameter
$\sigma$ for which the error goes to zero in the case of identical
oscillators, while the arrows labelled with ``2'' delimit the
interval of $\sigma$ for which the error goes to zero when the
oscillators have slightly detuned parameters.} \label{fgr:Erro}
\end{figure}

The new stability interval ${I^D_{st}=(\nu_1^D,\nu_2^D)}$ for the
selected intensity $D$ of noise may be found in the same way as it
has been described above by means of calculating the master
stability function $\Lambda^D(\nu)$. We will show that, under such
an assumption, the increase of the noise intensity $D$ leads the
boundaries $\nu_1^D$ and $\nu_2^D$ of the stability interval
$I^D_{st}$ to converge to asymptotic values $\nu_1^*$ and
$\nu_2^*$, respectively. These points $\nu_{1,2}^*$ are analogous
to the coupling strength value $\sigma_2$ in the case of two
coupled chaotic oscillators and determine the stability interval
$I_{st}^*$ for the considered network  of slightly non identical
elements.

To illustrate the proposed approach let us consider a network of
coupled R\"ossler systems. The dynamics of such network is
described by equation~(\ref{eq:Network}) with
${\mathbf{x}_i=(x_i,y_i,z_i)}$, $\mathbf{g}_i=(\omega_i)$,
$\mathbf{F}(\mathbf{x}_i,\mathbf{g}_i)=\mathbf{F}(\mathbf{x}_i,\omega_i)=[-\omega_iy_i-z_i,
\omega_ix_i+0.165y_i, 0.2+z_i(x_i-10)]$,
$\mathbf{H}[\mathbf{x}]=x$:
\begin{equation}
\begin{array}{l}
\dot x_i=-\omega_iy_i-z_i-\sigma \sum_{j=1}^NG_{ij}x_j,\\
\dot y_i=\omega_ix_i+0.165y_i,\\
\dot z_i=0.2+z_i(x_i-10).
\end{array}
\label{eq:ResslerNetwork}
\end{equation}
The mismatch in the parameters here corresponds to a detuning in
the natural frequencies $\omega_i$ of the oscillators, that are
supposed to be distributed randomly with a mean value
$\bar{\omega}=\langle\omega_i\rangle=1$ and a small dispersion
$\Delta\omega \cong 0.1 $.

For the case of identical R\"ossler oscillators (i.e. assuming all
oscillators to have the same natural frequency $\bar{\omega}=1$),
it exists a finite range of values for the parameter $\nu$ (that
we will call the stability interval $I_{st}$) for which the master
stability function is negative \cite{Pecora:1998_CS}.

To take into account the small difference in the frequencies of
the coupled oscillators, equations (\ref{eq:StabNetworkEquation})
and (\ref{eq:NoiseEvolutionEquation}) are instead  used to
calculate the stability interval $I_{st}^D$ as discussed above. To
calculate the master stability function $\Lambda^D(\nu)$
characterizing the property for synchronization of the network
with slightly detuned elements, we made use of a random process
$\mathbf{\xi}(t)$ distributed uniformly over the interval
$(-1.0;1.0)$. To integrate equation
(\ref{eq:NoiseEvolutionEquation}) the one-step method has been
applied~\cite{Garcia-Ojalvo:1999_NoiseBook} with the time step
$\Delta t=10^{-6}$.

The fragments of the dependence $\Lambda^D(\nu)$ around the
boundary points $\nu_{1,2}$ are shown in
Fig.~\ref{fgr:MSFIDetuned}. One can see that the influence of
noise in (\ref{eq:NoiseEvolutionEquation}) results in the shift of
the boundary points $\nu_{1,2}^D$ and in the consequent reduction
of the stability interval $I_{st}^D$. Therefore, the range of the
coupling strength value $\sigma$ corresponding to the synchronous
dynamics of the network of elements with slightly different values
of parameters is less in comparison with the analogous network
consisting of the identical elements.

The dependence of the normalized length $L_{st}^D/L_{st}$ (where
${L^D_{st}=\nu_2^D-\nu_1^D}$ and ${L_{st}=\nu_2-\nu_1}$,
respectively) of the stability interval $I^D_{st}$ on the intensity
of noise $D$ is shown in Fig.~\ref{fgr:StabInt}. One can see that,
under the increase of the noise intensity $D$, the length of the
stability interval $L^D_{st}$ converges to the value
$L_{st}^{\alkor{D}*}$ which does not depend practically on
$D$-value. At the same time, the boundaries $\nu_1^D$ and $\nu_2^D$
of the stability interval $I_{st}^D$ converge to the points
$\nu_1^*$ and $\nu_2^*$, respectively. Therefore, the obtained
interval $I_{st}^*$ is the region of parameter $\nu$-values
corresponding the stable synchronized behavior of the considered
network consisting of elements with slightly different parameter
values. It is important to note, that the stability interval
$I_{st}^*$ is found when the noise intensity $D$ is increased
step-by-step. At the same time, if the $D$-value used in
(\ref{eq:NoiseEvolutionEquation}) is too large (e.g., the noise
intensity is comparable to the amplitude of oscillations), the
dynamics of oscillator may be destroyed completely by noise, and, as
a result, the boundary points of the stability region $I_{st}^*$
will not be detected correctly. In other words, there is a range of
the reasonable values of the noise intensity $D$ corresponding to
the behavior of the network with slightly detuned elements.

\begin{figure}[tb]
\centerline{\includegraphics*[scale=0.4]{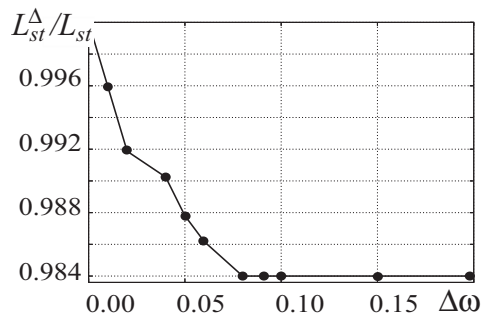}} \caption{ The
length $L_{st}^\Delta$ of the stability interval $I_{st}^\Delta$
{\it vs.} the maximum deviation of the control parameter value
$\Delta\omega$ for the network of R\"ossler oscillators with
slightly detuned parameters. The value $L^\Delta_{st}$ is normalized
to that of the length ($L_{st}$) of the stability interval $I_{st}$
for the network consisting of the identical R\"ossler oscillators.}
\label{fgr:StabInt_delta}
\end{figure}

In order to show that the approximate solution of the master
stability formalism is valid already for networks of relatively
small size, a direct numerical simulation of Eq.
(\ref{eq:ResslerNetwork}) has been carried out for various
different values of the control parameters. This calculation
allows to find the boundary of the stability of the synchronous
regime directly and to compare them with the analogous ones
obtained before (based on
Equations~(\ref{eq:StabNetworkEquation}),
(\ref{eq:NoiseEvolutionEquation})). We consider here a network of
$N=5$ non identical R\"ossler systems with $\Delta \omega_i=0.07$.
The coupling matrix $G$ is selected as
\begin{equation}
\left(
\begin{array}{ccccc}
-2& 0 & 0 & 1 & 1\\
0 & -3 & 1 &  1 & 1\\
0 & 1  & -3 &  1 &  1\\
1&  0 & 0 & -1 & 0 \\
1 & 1 & 1 & 0 & -3
\end{array}
\right),
\end{equation}
having as eigenvalues $\lambda_1=0.0$; $\lambda_2\approx-1.6$;
$\lambda_3\approx-2.0$; $\lambda_4\approx-4.0$;
$\lambda_5\approx-4.4$.

In the simulations, the appearance of a synchronous state can be
monitored by looking at the vanishing of the time average  (over a
window T) synchronization error
\begin{equation}
 \langle E \rangle =
\frac{1}{T(N-1)}\sum_{j>1}\int^{t+T}_t\|\mathbf{x}_j-\mathbf{x}_1\|dt'.
\label{eq:Error}
\end{equation}
In the present case, we adopt as vector norm
$\|\mathbf{x}\|=|x|+|y|+|z|$. Fig.~\ref{fgr:Erro} reports the
synchronization error versus $\sigma$ for a given network
topology. The Figure comparatively reports the case of identical
R\"ossler oscillators all with frequency $\omega=1$, and the case
of slightly non identical oscillators with frequencies distributed
around the same mean $\bar{\omega}=1 $ and with
$\Delta\omega=0.07$. One can see that the interval of $\sigma$ for
which the error goes to zero reduces in this case, in accordance
with the arguments extracted from the Master Stability Function
description.

A relevant issue concerns the possibility of establishing a
quantitative correspondence between the noise intensity $D$ in
equation~(\ref{eq:NoiseEvolutionEquation}) and the dispersion of
the control parameter values in~(\ref{eq:ResslerNetwork}). To
clarify this point, we consider the ratio $L_{st}^D/L_{st}$
between the length of the stability interval for non identical
elements in the presence of noise and the same length for the case
of identical systems. For the network of elements with the
slightly nonidentical control parameters
~(\ref{eq:ResslerNetwork}), the stability interval $I^\Delta_{st}$
may be defined as the coupling strength range where the
synchronization error~(\ref{eq:Error}) vanishes.

\begin{figure}[tb]
\centerline{\includegraphics*[scale=0.4]{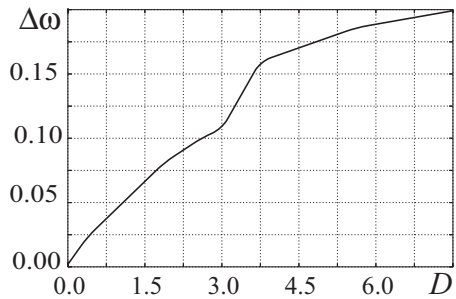}} \caption{ The
noise intensity $D$ {\it vs.} the control parameter deviation
$\Delta\omega$.} \label{fgr:delta_noise}
\end{figure}

The dependence of the normalized length $L_{st}^\Delta/L_{st}$ of
the stability interval $I^\Delta_{st}$ with the value of the
maximal deviation $\Delta\omega$ is shown in
Fig.~\ref{fgr:StabInt_delta}. By comparison with
Fig.~\ref{fgr:StabInt}, it is apparent how the curve in Fig.
~\ref{fgr:StabInt_delta} is in excellent agreement with its
analogue depicting the dependence of the normalized length
$L_{st}^D/L_{st}$ on the noise intensity $D$. Notice that, as the
the nonidentity of the network elements ($\Delta\omega$)
increases, the length $L_{st}^\Delta$ strives for its asymptotic
value $L_{st}^{\Delta *}$, which now does not depend on the
specific $\Delta\omega$-value. Moreover, this limit value
$L_{st}^{\Delta *}$ is in the good concordance with $L_{st}^{D*}$.
Finally, in Fig.~\ref{fgr:delta_noise} the relationship between
the noise intensity $D$ and the control parameter deviation
$\Delta\omega$ is reported, showing that the network of
non-identical oscillators can be suitably modelled by a noise
addition to the synchronization manifold characterizing the
evolution of the corresponding network of identical units.

In conclusion, we have estimated the property for synchronization
of networks consisting of equal elements with slightly different
control parameter values. This study may be considered as the
extension of the already known method of analysis of the behavior
of the networks of identical elements.

This work has been supported by Russian Foundation of Basic Research
(projects 05--02--16273, 06--02--16451, 06--02--81013). S.B.
acknowledges support from the Horovitz Center for Complexity. We
also thank the ``Dynasty'' Foundation.

\end{document}